\documentclass{80SA}
\usepackage{times}

\newcommand{\vecc}[1]{\mbox{\boldmath $#1$}}

\title{Upper critical field in superconductors near ferromagnetic
quantum critical points; UCoGe}

\author{Yasuhiro Tada\thanks{E-mail address: tada@scphys.kyoto-u.ac.jp},
Norio Kawakami, and Satoshi Fujimoto
}
\inst{Department of Physics, Kyoto University, Kyoto 606-8502, Japan
}
\abst{We study the strong-coupling superconductivity near 
ferromagnetic quantum critical points, mainly focusing on
the upper critical fields $H_{c2}$. 
Based on our simple model calculations,
we discuss experimentally observed unusual behaviors of $H_{c2}$ in
a recently discovered ferromagnetic superconductor UCoGe.
Especially,
the large anisotropy between $H_{c2}\parallel a$-axis and $H_{c2}\parallel
c$-axis, and the strong-coupling behaviors in $H_{c2}^{\parallel a}$ 
are investigated.
We also examine effects of non-analytic corrections
in the spin susceptibility on the superconductivity, which can arise
from effective long range interactions due to particle-hole
excitations.
}

\kword{superconductivity, upper critical field, 
ferromagnetism, quantum criticality,
non-analytic corrections}

\begin{document}
\maketitle

\section{Introduction}
Recently, a new ferromagnetic superconductor UCoGe was
discovered and 
has been attracting much interest \cite{pap:Huy07,pap:Huy08,
pap:Hassinger08,pap:Slooten09,pap:Aoki09,pap:Gasparini10,
pap:Deguchi10,pap:Ihara}.
It is considered to be located near a ferromagnetic
quantum critical point (QCP), and the related phenomena
are extensively studied.
In this compound, the superconductivity coexists with 
ferromagnetism in some pressure range,
and the pairing state is expected to be spin triplet with 
equal-spin pairing along the easy axis
of the magnetization, $c$-axis.
The superconductivity extends to the paramagnetic phase and 
the superconducting transition temperature changes smoothly
over the two phases, which suggests that the superconductivity
is mediated by the Ising ferromagnetic spin 
fluctuations \cite{pap:Monthoux99,pap:Wang01,pap:Fujimoto04}.

Intensive experimental studies reveal interesting
behaviors of the upper critical field $H_{c2}$ which would be
related to quantum criticality\cite{pap:Huy08,pap:Slooten09,pap:Aoki09}.
Naively, it may be expected that
the anisotropy in the 
upper critical field $H_{c2}$ in a superconductor with equal-spin 
pairing obeys
$H_{c2}^{\parallel c} > H_{c2}^{\perp c}$,
because there is no Pauli depairing effect for $c$-axis.
In the coexistence region of the ferromagnetism and the superconductivity
in UCoGe, however, the observed $H_{c2}$ is
$H_{c2}^{\perp c} \gg H_{c2}^{\parallel c}$,
which is completely different from the naive expectation.
Recently, it was pointed out that, for the superconductivity
coexisting with the ferromagnetism which has a sufficiently
large magnetization, the Pauli depairing effect for
a field perpendicular to the magnetization 
does not destroy the superconductivity and
the orbital depairing effect would be essential there \cite{pap:Mineev10}.
Although this explains a very important part of the interesting behaviors 
of $H_{c2}$ in UCoGe, there remain unresolved problems
which would be beyond the weak-coupling theory.
In addition to the unusual anisotropy, the behaviors of $H_{c2}\parallel 
\hat{a},\hat{b}$ are surprising \cite{pap:Slooten09,pap:Aoki09}.
$H_{c2}^{\parallel a}$'s are huge exceeding 10 (T) while the 
transition temperature $T_{sc}$ at zero field is 0.8 (K)
and have upward curvatures, which would be considered as a
characteristic property of the strong-coupling superconductivity.
The huge $H_{c2}$ with upward curvatures are also observed in
noncentrosymmetric heavy fermion superconductors 
Ce(Rh/Ir)Si$_3$ \cite{pap:Kimura07,pap:Settai08},
which can be understood as an interplay of lack of inversion symmetry
and quantum criticality \cite{pap:Tada}.
For $H_{c2}^{\parallel b}$ in UCoGe, $H_{c2}$ shows an S-shaped
curve at high fields 
and the similarity to that in URhGe was pointed out \cite{pap:Aoki09}.
These behaviors would be related to the ferromagnetic quantum criticality
in UCoGe.

Apart from experiments, there have been new theoretical progress
on ferromagnetic quantum criticality \cite{pap:Belitz,
pap:Duine05,pap:Conduit09,pap:Chubukov}.
It is argued that
the criticality can be affected by effective long range interactions
due to particle-hole excitations which lead to the so-called
non-analytic corrections.
If the superconductivity is mediated by ferromagnetic spin
fluctuations, it could be also affected by them.

In this study, we examine 
the strong-coupling superconductivity near ferromagnetic QCPs
having UCoGe in mind.
We restrict our calculations to the paramagnetic states and
neglect the Pauli depairing effect.
Under these conditions, we show that
$H_{c2}^{\parallel a} \gg H_{c2}^{\parallel c}$
can hold in some parameter range.
We perform the same calculations for two cases,
the case of the point node symmetry
and the line node symmetry, and conclude that
the point node symmetry is a promising candidate for the
superconductivity realized in UCoGe.
We also study the effects of non-analytic corrections
in the spin fluctuations on the superconductivity.

\section{Calculations and Results}
In this study, we use a very simple model to investigate
the superconductivity near the ferromagnetic 
QCP \cite{pap:Monthoux99,pap:Wang01,pap:Fujimoto04}.
The action is
\begin{eqnarray}
S&=&\sum_k c^{\dagger}_{k}[-i\omega_n+\varepsilon_k]c_{k}
-\sum_q \frac{2g^2}{3}\chi(q)S^z_qS^z_{-q},\\
\chi(q)&=&\frac{\chi_0}{\delta +c_nq^2\ln q +q^2 +|\Omega_n|/(vq)}
\label{eq:susceptibility}
\end{eqnarray}
where $k=(i\omega_n,\vecc{k})$, $c_{k}=(c_{k\uparrow},c_{k\downarrow})^t$,
$S_q^z=\sum_kc_{k+q}^{\dagger}(\sigma^z/2)c_k$,
$\varepsilon_k=-2t\sum_{i=a,b,c}\cos k_i-\mu$, and
the filling is fixed at $n=0.15$ for which the Fermi surface
is almost a sphere. We have taken the energy unit $t=1$ and the
length unit $a=1$.
The spin fluctuations are of Ising-type according to
the NMR experiments \cite{pap:Ihara} and the resulting
pressure-temperature phase diagram can be consistent with
the experiments \cite{pap:Monthoux99,pap:Wang01,pap:Fujimoto04}. 
The criticality is characterized by $\delta =\delta_0(T+\theta)
+c_h(\mu_BH_z)^2$, where we have just assumed the mean-field-like temperature
dependence.
Here, we fix $\delta_0=4$ and $v=4$.
The important point is that the spin fluctuations
depend on the applied magnetic field along the $c$-axis $H_z$, 
as suggested by 
the resistivity measurements \cite{pap:Aoki09} and 
the NMR experiments \cite{pap:Ihara}.
The non-analytic correction is incorporated with the
coefficient $c_n$ which can arise from the 
effective long range interactions due to particle-hole 
excitations \cite{pap:Belitz,
pap:Duine05,pap:Conduit09,pap:Chubukov}.
If we define the effective mass of the fluctuating modes as
$\delta_{\rm eff}\equiv \delta +{\rm min}[c_nq^2\ln q +q^2]$,
$\delta_{\rm eff}$ becomes smaller than $\delta$ for $c_n>0$, 
which means that the spin fluctuations
are enhanced by the non-analytic correction.
Here, we just put a remark on how actually the non-analytic
correction can appear.
In a fully SU(2) symmetric one-band model,
all the scattering processes with momentum transfer
$q\sim 0$ cancel out and cannot contribute to 
the spin susceptibility
because of the loop cancellation theorem \cite{pap:Metzner98}.
For such a case, the non-analytic correction can arise
from the scattering processes which include at least one 
scattering with $q\sim 2k_F$, where $k_F$ is the Fermi wavenumber.
On the other hand, the system with Ising-like symmetry,
there would be no strong reason by which $q\sim 0$
scattering processes completely cancel out.
This issue would be discussed in detail elsewhere.

To study the superconductivity focusing on 
the orbital depairing effect, we solve 
the Eliashberg equation without the Pauli depairing effect.
The linearized Eliashberg equation reads,
\begin{eqnarray}
\Delta_{\sigma \sigma}(k)&=&-\frac{T}{N}
\sum_{k^\prime}V(k,k^{\prime})
{\mathcal G}(k^{\prime})
\Delta_{\sigma \sigma}(k^{\prime}),\\
{\mathcal G}(k)
&=&\frac{1}{2}[\langle \phi_0|G(k+\Pi)
G(-k)|\phi_0\rangle \nonumber \\
&&+\langle \phi_0|G(k)
G(-k+\Pi)|\phi_0 \rangle ]
\end{eqnarray}
where $\Pi=(0,-i\nabla_R-2e\vecc{A}(\vecc{R}))$ with 
$\nabla \times \vecc{A}=(H,0,0)$ or
$(0,0,H)$, and
$\vecc{R}$ is the center of mass coordinate of the Cooper pair.
$\phi_0$ is the lowest level Landau function
and $G$ is the Green's function with the
selfenergy evaluated at the lowest order in $g^2\chi_0$,
$\Sigma (k)=(T/3N)\sum_qg^2\chi(q)G_0(k+q)$,
where $G_0$ is the non-interacting Green's function.
The pairing interaction is also calculated at the lowest order,
$V(k,k^{\prime})=-(1/6)g^2\chi(k-k^{\prime})
+(1/6)g^2\chi(k+k^{\prime}).$
The calculations are similar to those in the previous study \cite{pap:Tada}.
We fix $t=50$ (K) and $a=4.0$ (\AA) in this study,
which corresponds to the effective mass of the cyclotron motion
of the orbital depairing $m_{\rm eff}\equiv \hbar^2/(ta^2)\simeq 440
\times$(bare electron mass).

\begin{figure}[htb]%
\vspace{-0.5cm}
\begin{tabular}{ll}
      \resizebox{38mm}{30mm}{\includegraphics{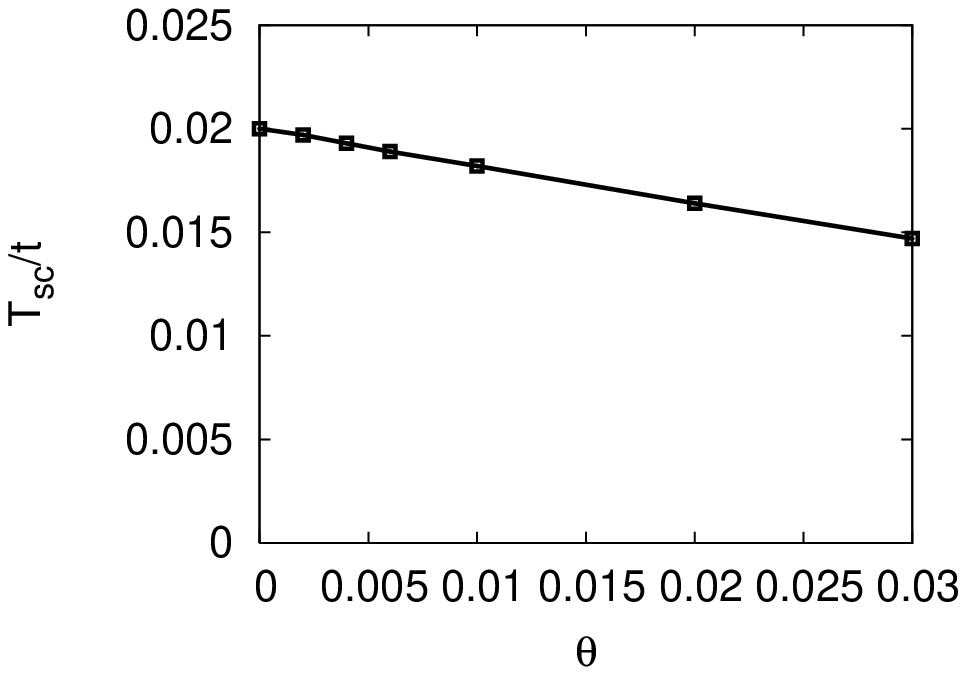}} &
      \resizebox{38mm}{30mm}{\includegraphics{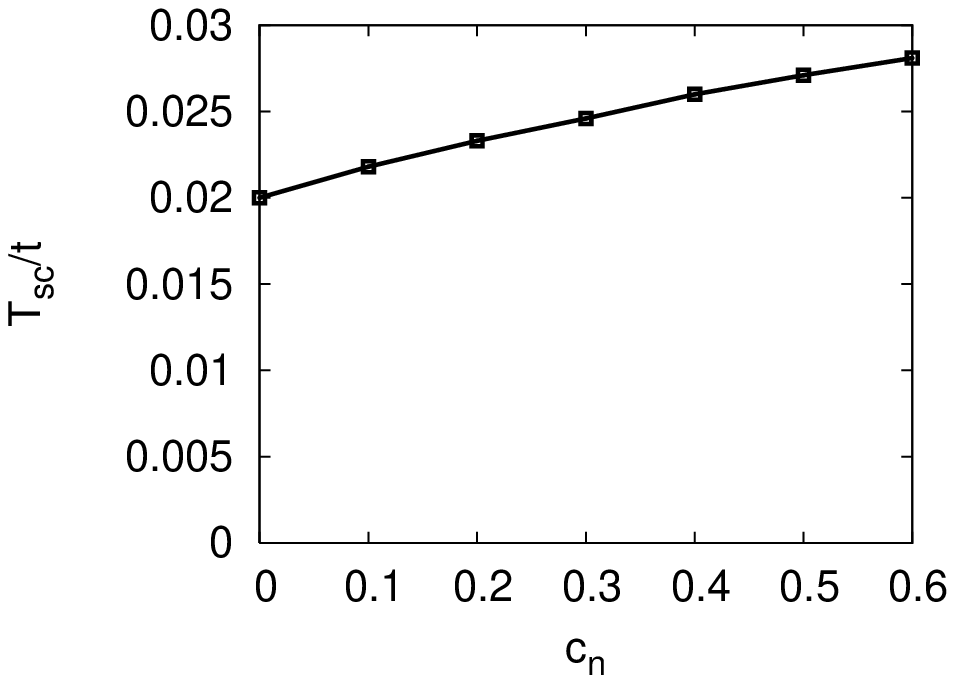}} \\
\end{tabular}
\caption{%
(Left panel) Transition temperature $T_{sc}$ as a
function of $\theta$ for $c_n=0$.
(Right panel) $T_{sc}$ as a function of $c_n$ for $\theta=0$.
}
\label{fig:T_sc}
\vspace{-0.5cm}
\end{figure}
First, we study the effects of the non-analytic
correction in the spin fluctuations on 
the superconductivity at zero field.
At zero field, the transition temperature for the gap 
function $d=(\sin k_a,\sin k_b,0)$ with the point nodes and the gap function
$d=(\sin k_c,0,0)$ with the horizontal node are degenerate.
In this study, the coupling constant is fixed as $g^2\chi_0=112.6t$
which is large
so that we have rather high transition temperatures within our model.
For this value of $g^2\chi_0$, the mass enhancement factor due to
the ferromagnetic spin fluctuations is $z\sim 0.8$.
The left panel of Fig. \ref{fig:T_sc}
shows the $\theta$-dependence of the transition temperature $T_{sc}$
which corresponds to the pressure-temperature
phase diagram of the paramagnetic side in the experiment and
$\theta =0$ corresponds to the QCP.
The transition temperature $T_{sc}$ is highest at $\theta=0$ and
is decreased as $\theta$
increases, which is consistent with the experimental phase
diagram of UCoGe \cite{pap:Hassinger08,pap:Slooten09}.
In the right panel of Fig. \ref{fig:T_sc}, 
we show $T_{sc}$ as a function of the coefficient of 
the non-analytic correction $c_n$.
We see that $T_{sc}$ is monotonically enhanced as $c_{n}$ increases.
This is because the effective mass of the criticality $\delta_{\rm eff}$
becomes small with finite $c_{n}>0$. 
This shows that the enhanced criticality by the non-analytic
correction favors the superconductivity
in the case of Ising-type spin fluctuations. 

Next,
we move to the effects of a magnetic field on the superconductivity
and fix $\theta=0$.
In the presence of the magnetic field, we use the same model,
eq.(\ref{eq:susceptibility}), 
although there might be other non-analytic corrections 
related to the magnetic field \cite{pap:Chubukov,pap:Misawa}.
Even if we neglect Pauli depairing effects, the degeneracy of 
the symmetries of the gap functions is lifted by $H$ and
the anisotropy of $H_{c2}$ depends on the positions of the 
gap nodes \cite{pap:Scharnberg85}.
This is simply explained as follows.
The effective velocity for
the cyclotron motion of the orbital depairing effect can be
of the form of $\tilde{\vecc{v}}_k=\varphi(\vecc{k})\vecc{v}_k$ for 
the basis function $\varphi$ corresponding to the superconducting gap
symmetry.
When the magnetic field is parallel to the $c$-axis,
the cyclotron motion can enjoy the nodes for the case with
horizontal line nodes where $\tilde{\vecc{v}}_k=0$, resulting in a large
orbital limiting field.
The cyclotron motion cannot do so for $H\perp \hat{c}$, which leads to
the anisotropy of the orbital limiting field 
$H_{c2}^{\parallel c}>H_{c2}^{\perp c}$.
On the other hand, under $H\parallel \hat{c}$,
$\tilde{v}_k$ cannot be zero for the point node
gap function except for the poles of the Fermi surface, $k_a=k_b=0$.
Therefore, for point nodes at the poles of the Fermi surface, 
the order is turned over,
$H_{c2}^{\parallel c}<H_{c2}^{\perp c}$.

\begin{figure}[htb]%
\vspace{-0.5cm}
\begin{tabular}{ll}
      \resizebox{38mm}{30mm}{\includegraphics{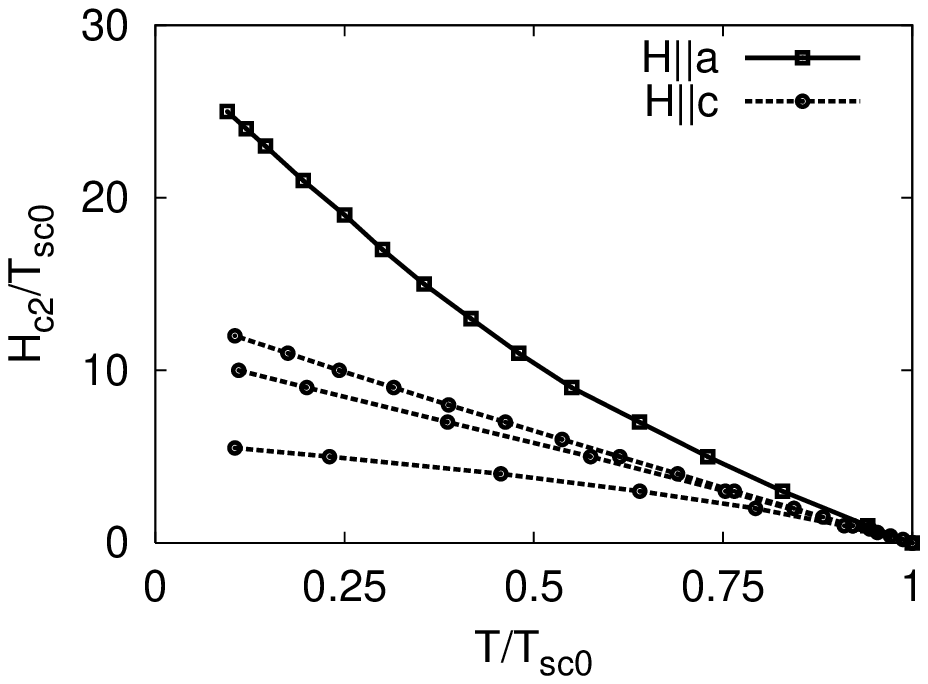}} &
      \resizebox{38mm}{30mm}{\includegraphics{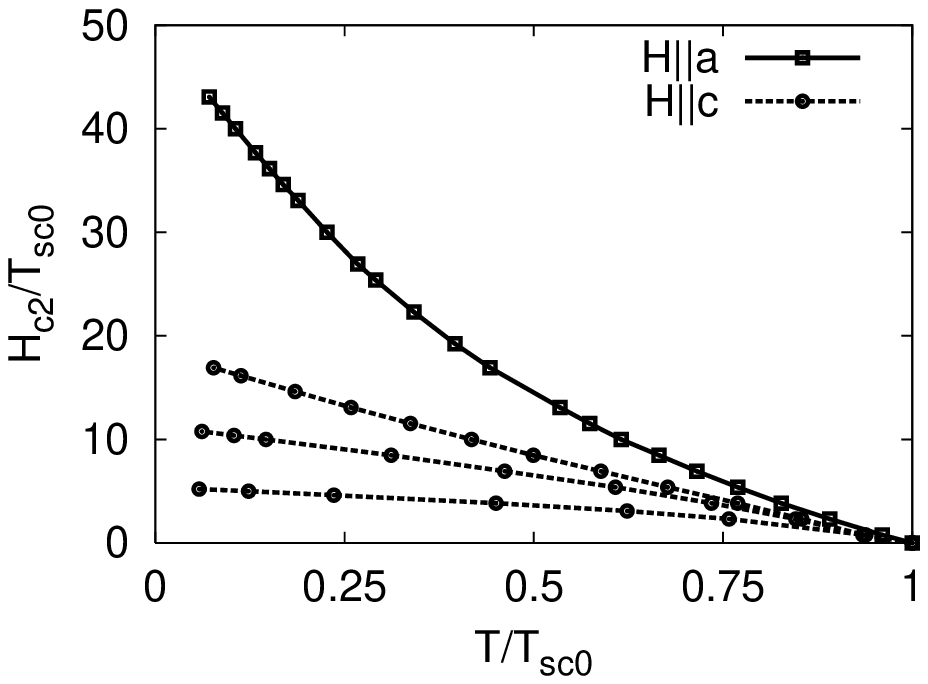}} \\
\end{tabular}
\caption{%
Temperature v.s.
upper critical fields $H_{c2}$
for $c_n=0$ (left panel) and $c_n=0.4$ (right panel)
for $d=(\sin k_a,\sin k_b,0)$.
The solid curves are for $H_{c2}\parallel a$-axis, and
the dotted curves are for $H_{c2}\parallel c$-axis with 
$c_h=0, c_h=1.0, c_h=10.0$ from the top to the bottom,
respectively.
}
\label{fig:pt}
\vspace{-0.5cm}
\end{figure}
In Fig. \ref{fig:pt}, we show the upper critical field for several values
of $c_h$ for the point node case for which $d$-vector
is $d=(\sin k_a,\sin k_b,0)$.
$H_{c2}$ is normalized by the transition temperature at zero field
$T_{sc0}$ which is a measure of the Pauli limiting field for the
usual $s$-wave superconductivity.
The solid curve is for $H\parallel \hat{a}$ and dotted curves are
for $H\parallel \hat{c}$.
We see that, even if $c_h=0$, $H_{c2}^{\parallel a}>
H_{c2}^{\parallel c}$ holds because of the point nodes.
When the Ising spin fluctuations are suppressed by the magnetic
field $H\parallel \hat{c}$ as suggested by the 
experiments \cite{pap:Aoki09,pap:Ihara}, 
$H_{c2}^{\parallel c}$ is reduced and 
it has only moderate temperature dependence for sufficiently large $c_h$.
If there exists strong suppression of the spin fluctuations,
$H_{c2}$ can have strong anisotropy $H_{c2}^{\parallel a}
\gg H_{c2}^{\parallel c}$.
Note that, for the Ising anisotropic case,
the spin fluctuations are robust against in-plane fields,
resulting in that $H_{c2}^{\parallel a}$ remains large
with the strong-coupling behaviors.
The non-analytic correction can make the result more
drastic and the calculated anisotropy is quite strong as
seen in the right panel of Fig. \ref{fig:pt}.

The strong enhancement of the calculated $H_{c2}^{\parallel a}$ is
understood as a result of an increasing pairing interaction
and a decreasing depairing effect of the spin fluctuations
at low temperatures.
The physical origin is the same as that of the huge $H_{c2}^{\parallel c}$
in noncentrosymmetric heavy fermion superconductors 
Ce(Rh/Ir)Si$_3$ \cite{pap:Kimura07,pap:Settai08,pap:Tada}.
Therefore, as pointed out in the previous work \cite{pap:Tada},
it can be said that
the strong enhancement in the orbital limiting field
near the quantum criticality has universal nature.
In this study, however, we have assumed that
the superconductivity is orbital limited, and
the suppression of the Pauli depairing effect \cite{pap:Mineev10}
would work well in the coexistence region of the ferromagnetism
and the superconductivity in UCoGe.
In the coexistence region, the mass of the criticality $\delta$
is replaced with the magnetization which has weak temperature dependence.
Whether such a 
mass can actually
lead to huge $H_{c2}^{\parallel a}$ without the assumption of
the orbital-limited superconductivity is an open issue
which should be addressed in the near future.

\begin{figure}[htb]%
\vspace{-0.5cm}
\begin{tabular}{ll}
      \resizebox{38mm}{30mm}{\includegraphics{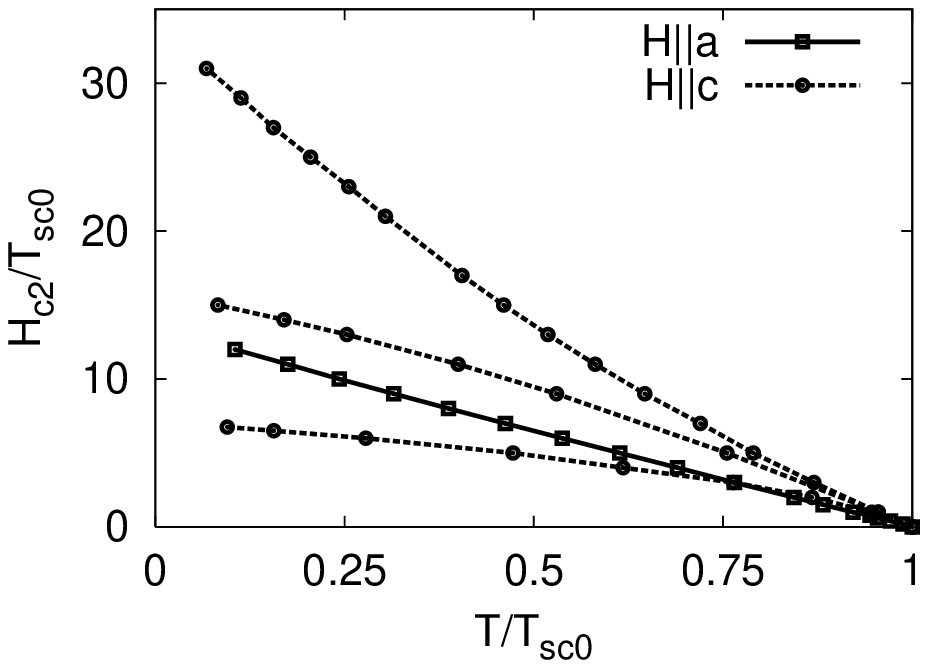}} &
      \resizebox{38mm}{30mm}{\includegraphics{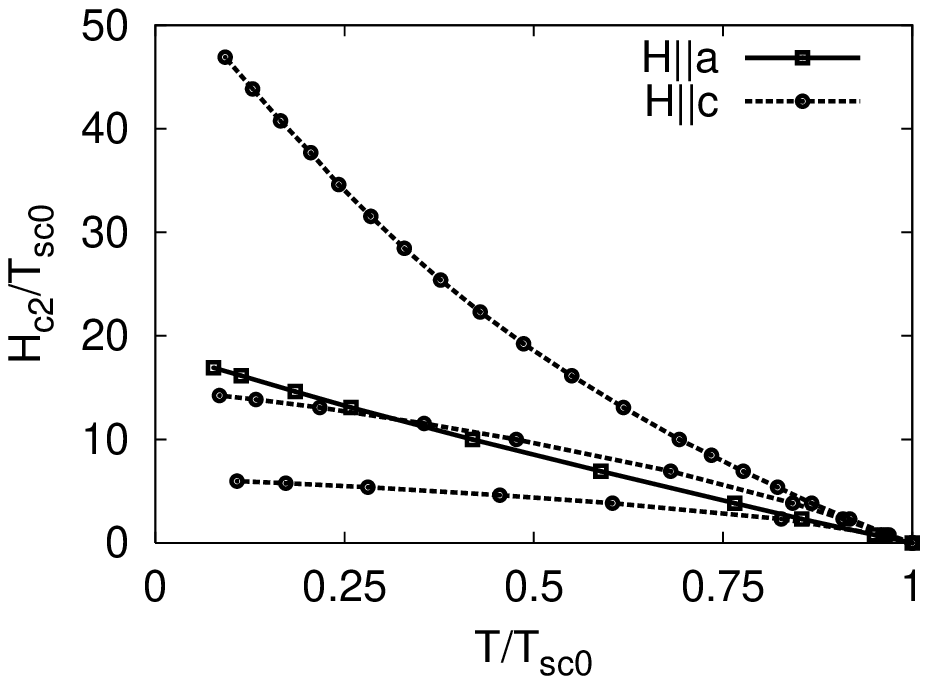}} \\
\end{tabular}
\caption{%
Temperature v.s.
upper critical fields $H_{c2}$
for $c_n=0$ (left panel) and $c_n=0.4$ (right panel)
for $d=(\sin k_c,0,0)$.
The solid curves are for $H_{c2}\parallel a$-axis, and
the dotted curves are for $H_{c2}\parallel c$-axis with 
$c_h=0, c_h=1.0, c_h=10.0$ from the top to the bottom,
respectively
}
\label{fig:line}
\vspace{-0.5cm}
\end{figure}

On the other hand, for the case of the gap function 
$d=(\sin k_c,0,0)$ with a horizontal line node, the relation
$H_{c2}^{\parallel a}<H_{c2}^{\parallel c}$ holds
when the suppression of the spin fluctuations is not so
strong as shown in Fig. \ref{fig:line}. 
For $H_{c2}$ to be $H_{c2}^{\parallel a}>H_{c2}^{\parallel c}$,
a quite large $c_h$ is needed, which cannot be usually realized.

From the above results, we conclude that a gap function with
point nodes is more promising than that with 
horizontal line nodes as a candidate for UCoGe in the coexistence region.
In UCoGe which has a orthorhombic structure,
possible gap functions for the coexistent states 
are classified by the magnetic point group $D_2(C_2^z)$ \cite{pap:Mineev}.
Among the possible symmetries, the most promising candidate is
the A$_1$ state or the A$_2$ state
for $H_{c2}$ to be consistent with the experiments
\cite{pap:Huy08,pap:Slooten09,pap:Aoki09}.

In summary, we have discussed the superconductivity near
the ferromagnetic QCP within a simple model for the Ising spin fluctuations.
The non-analytic correction can enhance the
superconducting transition temperature.
Under a magnetic field,
the positions of the nodes strongly affect the orbital cyclotron motion,
leading to the anisotropy in $H_{c2}$.
Taking into account the suppression of the Ising spin fluctuations
by $H\parallel \hat{c}$, we have shown that the anisotropy can be
$H_{c2}^{\parallel c}\ll H_{c2}^{\parallel a}$ for 
the point node case, and $H_{c2}^{\parallel a}$
has characteristic nature for strong-coupling superconductivity.
On the other hand, it is rather hard to reproduce 
such behaviors for the horizontal 
line node case.
These results suggest that the
superconductivity realized in UCoGe is in the A$_1$ state or
the A$_2$ state, both of 
which have no horizontal line nodes.

We thank K. Ishida, Y. Ihara, and D. Aoki for valuable discussions.
Numerical calculations are partially performed at the Yukawa Institute.
This work is partly
supported by the Grant-in-Aids for Scientific Research from MEXT
of Japan (Grant Nos. 19052003, 20102008, 21102510, and
21540359)
and the Grant-in-Aid for the Global COE Program
"The Next Generation of Physics, Spun from Universality and
Emergence."
N. K. is supported by JSPS through its FIRST program.
Y. T. is supported by JSPS Research Fellowships
for Young Scientists.

\end{document}